# Information-Driven Modeling of Biomolecular Complexes

Charlotte W. van Noort[#], Rodrigo V. Honorato[#], Alexandre M.J.J. Bonvin*

Bijvoet Centre for Biomolecular Research, Faculty of Science, Department of Chemistry, Utrecht University, Padualaan 8, 3584CH Utrecht, Netherland

# These authors contributed equally

* Corresponding author: a.m.j.j.bonvin@uu.nl

**Abstract**

Proteins play crucial roles in every cellular process by interacting with each other, with nucleic acids, metabolites, and other molecules. The resulting assemblies can be very large and intricate and pose challenges to experimental methods. In the current era of integrative modeling, it is often only by a combination of various experimental techniques and computations that 3D models of those molecular machines can be obtained. Among the various computational approaches available, molecular docking is often the method of choice when it comes to predicting 3D structures of complexes. Docking can generate particularly accurate models when taking into account the available information on the complex of interest. We review here the use of experimental and bioinformatics data in protein-protein docking, describing recent software developments and highlighting applications for the modeling of antibody-antigen complexes and membrane protein complexes, and the use of evolutionary and shape information.

**Keywords:** Macromolecular Docking; Integrative Modeling; Protein-Protein Interactions; Interaction Restraints; shape; co-evolution; membrane complexes; antibody-antigen complexes

# Introduction

Macromolecules such as proteins and nucleic acids are involved in most cellular functions responsible for maintaining life, performing their tasks in most cases by interacting with other molecules. Understanding these interactions is fundamental, not only to gain insight into the molecular machinery of living organisms but also to gather high quality information to drive innovation in, for example, protein engineering and drug design.

Theoretical approaches to the study of macromolecular interactions, such as docking, take full advantage of robust spatial search algorithms like rigid-body minimization, swarm optimization, and grid-based search methods (to name a few) to probe the interaction space of the molecular components of a complex. These algorithms can generate hundreds to tens of thousands of possible conformations, models, which must be scored and ranked [1]. The scoring and ranking of models are crucial steps in docking. Good scoring functions must be able to identify which models are valuable, representing near-native conformations, and should be further analyzed. This can be done in different ways: Classical approaches use a molecular engine to evaluate energetic components such as inter-molecular van der Waals, electrostatic, desolvation energies, etc. Alternatively, one might use statistical potentials derived from the analysis of known complexes. More recently the field has seen a rise in the use of machine learning techniques for scoring [2–6].

Over the last years, computational methods to study macromolecular interactions have been steadily incorporating different types of data to guide, filter, or validate their predictions [7–13]. The use of various types of information in macromolecular docking is commonly referred to as integrative modeling and has been a convergence point in the field, being implemented in most software under active development [14]. Such information can be used *a priori*, to guide the spatial search step, or *a posteriori*, to aid in the scoring of models (Figure 1). A perfect example is the Integrative Modeling Platform (IMP) [15], a renowned tool that can handle highly heterogenous information. IMP has been used to unveil the structure and functional anatomy of a nuclear pore complex [16], the 26S proteasome holocomplex [17], and the molecular architecture of the yeast spindle pole body core [18]. As in this review the main focus is on docking, IMP, which is formally not a docking platform, will not be discussed further.

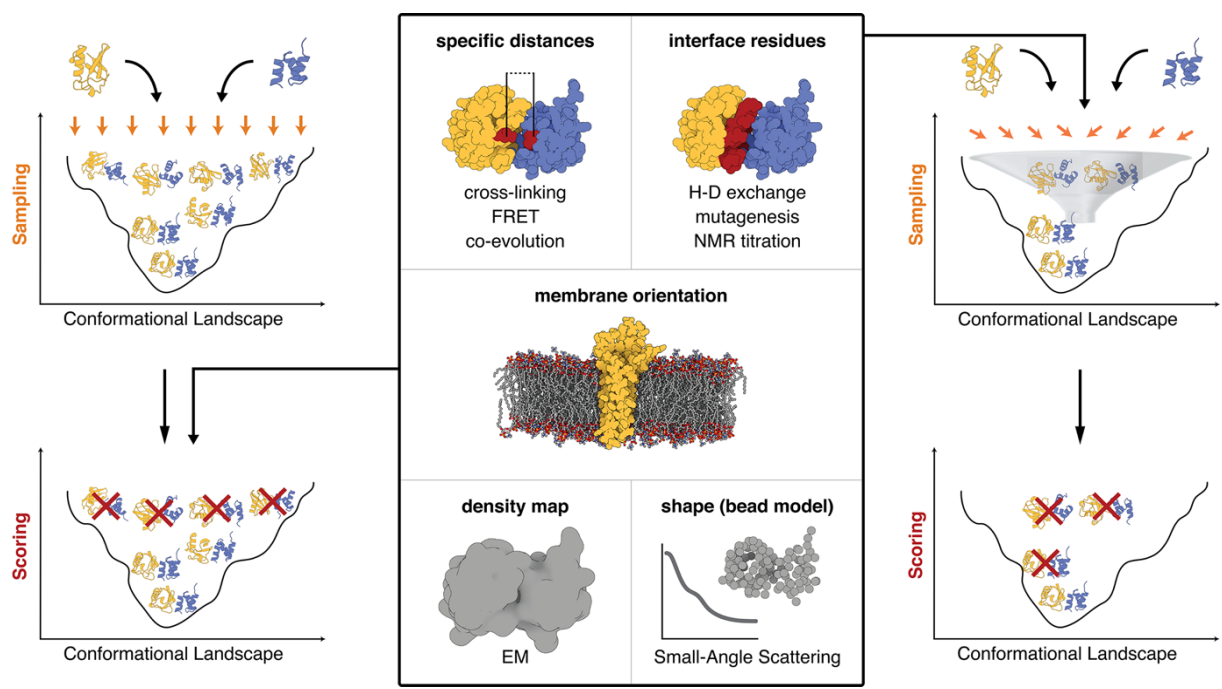

**Figure 1:** This figure illustrates the information-driven modeling of biomolecular comples, with in the central panel an illustration of various information sources, on the left a docking protocol that would only make use of the information in the filtering stage after sampling the interaction space, and on the right an information guided docking protocol that uses the data to bias the sampling and score the resulting models.

To drive molecular docking, data from a variety of experimental or computational sources can be used. To mention a few, hydrogen deuterium exchange experiments allow to identify regions of a target molecule that become protected from exchange upon binding, information which can be used to define the interface region [19]. Mutagenesis experiments can reveal if a residue is essential for the interaction, but give no specific information about its position in the interface or the contacts it makes [20]. Cross-linking experiments detected by mass spectrometry provide specific information between pairs of residues in the form of maximum distances, the length of which depends on the nature of the cross-linker reagent used [21,22]. Interaction between molecules can also be studied by measuring the Förster Resonance Energy Transfer (FRET) that occurs when two fluorescent-labeled proteins are in close proximity of each other [23]. FRET measurements can demonstrate the interaction between molecules *in situ*. Coupled with quantitative analysis they can provide valuable information for modeling protein structures and their complexes [23].

Interface and distance information from experimental methods can be incorporated into docking to make sure that the resulting models match the experimental information. In principle any method that can provide some kind of interface and/or distance information can benefit docking [1]. Since each technique has advantages, disadvantages, and limitations, most studies rely on information obtained with more than one experimental method. Whilst experimental data provide in principle higher quality information, access to experimental facilities and sample availability are often major limiting factors, especially for large-scale studies. In such cases, bioinformatics analysis can leverage large volumes of sequence information to yield valuable predictions about interfaces, as well as specific contacts. The latter can be extracted by statistical analysis of co-evolving residues in multiple sequence alignment [24].

In this review we focus on how different docking software use a variety of data in their predictions, as well as recent applications in the integrative modeling of biomolecular complexes, especially antibody-antigen and membrane complexes. We also discuss some recent developments in the use of evolutionary information and give an outlook on the use of shape information in biomolecular modeling.

**Docking Software**

HADDOCK (High Ambiguity Driven DOCKing) pioneered the use of experimental information in macromolecular docking [25]. It allows interface information to be entered as a set of active residues that have the highest probability of being part of the interaction interface and passive residues which are likely in the vicinity of the interface. This set of residues defines Ambiguous Interaction Restraints (AIR), associated with a maximum effective distance that draws the interfaces together without pre-defining their relative orientation. HADDOCK also supports the definition of specific distance restraints, e.g., from cross-linking MS experiments or co-evolution predictions. In the HADDOCK score used to rank the resulting models, a penalty is assigned for restraints that are not satisfied and combined with intermolecular energy terms.

Using the same format as for HADDOCK, a file containing restraint definitions can be submitted to ATTRACT [26]. ZDOCK accepts a list of contacting residues, which is used to filter rigid body docking solutions in which these are not near the other molecule [27]. This *a posteriori* filtering is also done by ClusPro, which accepts distance restraints that are used to select conformations that match the available data. The score used to rank the resulting models, however, does not include any restraint term [28]. pyDock [13], another rigid body docking method, offers *a posteriori* use of distance restraints by the pyDockRST module, which provides a score that combines the percentage of satisfied restraints with electrostatics and desolvation energies [29]. Distance restraints have also been implemented in the protein-peptide docking software CABS-dock and its web server [30,31].

One of the newest members in the family of integrative docking software is LightDock [32]. This software uses a swarm-based algorithm to distribute initial positions of the ligand relative to the receptor and is able to take into account interface information in different ways [7]. Given a set of interface residues on the receptor, the ligand swarms are positioned only around the defined interface region; each swarm is then optimized using particle swarm optimization. If the interface on the ligand side is also known, the molecule is rotated in relation to the receptor so that the specified residues face the receptor in the starting swarm orientations. The scoring of the models during the docking also reflects the interface information. The resulting conformations are filtered to include only those that are closest to the defined interface. Rather than assigning a penalty for unsatisfied restraints, in LightDock a bonus is defined based on the percentage of interface residues that are in contact with the binding partner.

The Exhaustive Rotational Search based Docking (EROS-DOCK) [33] is another docking software that uses information to avoid sampling the subspace that does not satisfy a given restraint. It also belongs to the very exclusive group of docking software that can handle both restraints and multi-body docking (i.e., the modeling of complexes consisting of more than two components), together with the pioneer HADDOCK and ATTRACT [25,34]. EROS-DOCK, in contrast to grid-based docking methods, applies a quaternion $\pi$-ball that represents the space of all possible Euler rotations. This $\pi$-ball representation is systematically explored with the objective of locating the global minimum of pairwise docking energies, avoiding steric clashes. Here restraints can be defined as amino acid or atom pairs together with their maximum separation distance. These restraints are used to build a constraint $\pi$-ball which is then used to identify poses that will never satisfy the restraints and thus should be discarded. The application of this spatial sampling methodology to both Protein-Protein Docking Benchmark v4 [35] and a self-made multi-body benchmark resulted in higher quality models with than without restraints [9•].

Several docking software also support the use of density- or shape-related information, e.g., from cryo-Electron Microscopy (cryo-EM) or Small-Angle X-ray Scattering (SAXS). These are discussed in a separate section further down. A non-comprehensive list of docking software and which type of information each of them can use is presented in Table 1.

**Table 1:** List of docking software mentioned in this review and the type of information they can use.

| Docking Software | Residue information[a] | Specific contacts[b] | Density/Shape/ Topological information[c] | Reference |
|---|:---:|:---:|:---:|:---:|
| ATTRACT | ● | ● | ● | [34,36,37] |
| ClusPro | ● | ● | ● | [28] |
| DOCK/PIERR |  | ● |  | [38] |
| EROS-DOCK | ● | ● |  | [9, 33] |
| FlexEM |  | ● | ● | [39] |
| FoXSDock |  |  | ● | [40,41] |
| HADDOCK | ● | ● | ● | [11,25,42,43] |
| InterEvDock |  | ● |  | [44,45] |
| JabberDock | ● |  | ● | [46] |
| LightDock | ● | ● | ● | [7,32,47] |
| Memdock | ● |  |  | [48] |
| pyDock |  | ● | ● | [13,49] |
| Rosetta | ● | ● | ● | [50–53] |
| ZDOCK | ● | ● |  | [27] |

a) Residue-level information about specific residues that are important for the interaction, but without specific information about the contact they make or their position within the interface
b) Contacts / distances between specific pairs of residues or atoms.
c) Density/shape information about the shape of the complex (e.g., from cryo-EM and SAXS) and topological information such as provided for example by the membrane for membrane-related complexes

## Recent Developments in Antibody-Antigen Docking

To investigate the possibilities and limitations of antigen-antibody docking, Ambrosetti *et al* recently compared the performance of ClusPro, HADDOCK, LightDock, and ZDOCK [54], four commonly used software that allow to make use of the knowledge about the antibody hypervariable loops. The software was tested in three setups: using information about the hypervariable loops but no information about the epitope, with a low resolution definition of the epitope, and with the real interface observed in experimental structures. HADDOCK outperformed the other software both in quality of generated models and success rate and a detailed protocol has been made available [55].

The RosettaAntibody and Rosetta SnugDock methods for antibody structure prediction and antibody–antigen docking have also been made more robust, with a simplified user interface, an expanded and automated template database, options to model single-domain antibodies with a more generalized kinematics engine, and also new loop modeling techniques

[50]. Very recently, an updated, extended and more diverse benchmark for antibody-antigen docking was published, including binding affinities. In relation to docking, authors compared ZDOCK, ClusPro and Rosetta SnugDock and highlighted the challenges posed by monoclonal antibodies, interactions with glycoproteins and camelid nanobodies [56•].

## Moving into the Membrane

The majority of docking software has been developed and benchmarked on soluble complexes. Membrane proteins (MP) and their complexes, which are involved to great extents in vital biological processes, have received so far rather limited attention. Many MPs act as receptors and are involved in signal transduction pathways. Understanding how such proteins interact with other macromolecules is therefore of great value for drug discovery. The fast development of molecular crystallography techniques such as *in situ* data collection, micro-crystallography, cryo-EM, and other state-of-the-art experimental techniques [57] have shed a new light on MPs [58].

In the meanwhile, macromolecular complexes involving MPs can now also be modelled by docking using experimental data and/or the information provided by the membrane itself. Several docking software have implemented specialized protocols for modeling membrane protein complexes, including DOCK/PIERR[38], Memdock[48], and RosettaMP [53]. In a recent publication, LightDock was combined with HADDOCK in a novel approach to model membrane-associated protein assemblies [47••]. This novel protocol proposes a way to study the interaction between a MP and a free ligand which is not bound to the membrane by using the "meta-information" of the membrane topology. The latter is taken from the MemProtMD database [59] which uses coarse-grained molecular dynamics simulations to produce a theoretical model of the membrane architecture around the protein. LightDock is able to leverage this "meta-information" by taking into account the simulated membrane topology, considering only the phosphate atoms and limiting the search space to search points that are outside the membrane. The membrane itself is also used to penalize models that would penetrate it.

While LightDock's membrane protocol is designed to dock soluble ligands to trans-membrane proteins, JabberDock recently introduced an approach for modeling trans-membrane dimers [46•]. This method uses surfaces derived from Spatial and Temporal Influence Density (STID) maps, which represent the dynamics and electrostatics of a protein based on a short Molecular Dynamics (MD) simulation. In the docking process, the surface complementarity of the two interaction partners is then maximized. For membrane proteins, the MD simulations are performed on the individual proteins embedded in a membrane, so that the properties of the TM region are captured in the resulting STID maps [46].

## Evolution to the rescue

A large portion of software that support the definition of restraints uses the classical definition of pairwise distances, which, although being a reductive interpretation of highly complex experiments such as cross-linking and mutagenesis, have been producing high quality models. Distance information to be used in docking can also be extracted from co-evolution analysis of pairs of proteins [60,61]. A docking software that focuses on the integration of evolutionary information is InterEvDock [44,45,62]. In a recent work, the InterEvDock group used the targets from the community-wide Critical Assessment of PRediction of Interactions (CAPRI) [63] rounds 38-45 to explore the extent to which evolutionary information can be used to model protein-protein complexes [62•]. By deriving recurrent interface features from homologous interfaces, applying techniques that were used for covariation-based folding, and by using template-based docking they were able to generate acceptable or better models in the top 5 predictions for 11 targets. Template-based approaches also fall into the category of evolutionary information since they are based around homology. These have been discussed elsewhere [64].

Not only can co-evolution analysis be used to derive distance restraints for the modeling of complexes, but it can also be applied to predict which proteins interact on a proteome-wide scale, as recently demonstrated on *E. coli* genome sequence data [65]. In this work, a logistic regression model was derived using a set of true positive interacting pairs with known structures as well as non-interacting pairs, based on yeast-two-hybrid experiments. The focus of the analysis was cell membrane protein interactions, which include approximately 1.25 million potential pairs. Using their EVcouplings framework [24] the authors could reveal 529 novel protein interactions and their interacting residues. The latter were then used in HADDOCK to model the predicted complexes.

Additionally, co-evolving mutations usually represent key residues involved in physical coupling and these can be determined from an analysis of multiple sequence alignments and provide valuable evolution-based information for docking [66]. The concept is not novel and was introduced already in the '90s by Valencia and co-workers [67], but recent tools like pydca [61] and various web servers such as, for example, EVcouplings [24], RaptorX ComplexContact [68] are simplifying their use. Many of those servers have participated to the CASP contact prediction experiment whose results have been discussed in the related CASP assessment papers [69].

**Use of shape information in macromolecular docking**

Although cryo-EM provides increasingly high-resolution electron density maps, these are not always sufficient to obtain a complete model of a biomolecular complex at an atomic level and this is even more true in the case of cryo-Electron Tomography (cryo-ET). The use of EM densities has been implemented in several rigid-body fitting methods, from the grid-based tool CoLoRes in Situs [70] to MultiFit [71]. FlexEM, which can be run by the MODELLER software [39], combines rigid-body fitting to a cryo-EM density with refinement, where parts of the structure are kept rigid, for example the secondary structure elements [72]. FlexEM can also incorporate distance restraints like those described above. These methods usually fit one protein at a time into the density, thus neglecting the intermolecular interactions in this process.

Docking methods that actually account for the interface of a complex while guiding the modeling with cryo-EM data include ATTRACT-EM [36] and HADDOCK [11,42]. In ATTRACT_EM, the resolution of the density map is reduced for the initial fitting of the components, after which the top models are refined in the original map [36]. In HADDOCK, centroids are first placed within the density map and (ambiguous) distance restraints are used to draw each molecule into its predicted position within the density. During the following refinement steps, the molecules are then restrained by the EM density itself. These programs can simultaneously apply classical distance restraints in addition to cryo-EM restraints.

Lower resolution shape information can be provided by Small Angle Scattering (SAS) methods such as Small-Angle X-ray or Neutron Scattering (SAXS/SANS) [73]. Scattering data from such experiments can be used in several docking methods, including SASREF [74], pyDockSAXS [49,75], IMP's FoXSDock [40,41], [37] and HADDOCK [76]. All of these methods use SAXS data to filter models, selecting them based either directly on the fit ($\chi$) of their theoretical scattering curve to the experimental one, or by integrating the $\chi$ value in a more generalized score. ClusPro and RosettaDock$_{SAXS}$ filter models in a similar way, but disregard $\chi$ in the final scores (and thus ranking) that they return [51,52]. Some can also incorporate a radius of gyration (derived from SAXS data) restraint, as implemented in HADDOCK.

SAXS scattering curves are also commonly translated into bead representations consisting of a set of dummy atoms to visualize the shape of a molecule or complex, with tools such as those available for example from the ATSAS software suite[77,78]. To our knowledge, the only method that makes use of such bead representations for protein-protein docking is ATTRACT-SAXS, where the search space is constrained by an atom density mask derived from a bead model [37]. Analogous applications have been reported in the field of small-molecule data, where the binding pocket may be defined as a set of 3D points [73]. This information can be used to define restraints that guide the small molecule to the correct position. Such bead models could provide a very versatile manner of representing a variety of experimental data such as SAS, low to medium resolution EM data, or any kind of volumetric data, and we expect that they will find their way into macromolecular docking in the near future. Note that shape information has been used in IMP as low-resolution representations of components for which no high-resolution 3D structures are available [79], as well as in LightDock, which uses a very simplified bead representation on the membrane (see above).

## Conclusions

An increasing variety of both experimental and predicted information can be used in the modeling protein-protein complexes. While classical (ambiguous) distance restraints work well in many scenarios, other creative ways of accounting for data on the interface as well as other characteristics of the complex are being developed. Protocols harvesting specific information for a given type of complex, such as those involving membrane-embedded proteins and antibody-antigen interactions, allow to generate models with increasing accuracy. With the explosion in genomic data, information extracted from sequence information is now increasingly used in several evolution-centered docking approaches. And finally, a growing number of methods now allow for shape-based information to be incorporated to ensure that the global shape of the generated models matches the experimentally derived shapes. All these developments clearly underscore the continuous increase in the use of information to drive the modeling of biomolecular complexes in the current era of integrative structural biology.

## Acknowledgments

The authors acknowledge the financial support from the European Union Horizon 2020 projects BioExcel (675728, 823830).

## References

1. Koukos PI, Bonvin AMJJ: **Integrative modelling of biomolecular complexes**. *J Mol Biol* 2019, **432**:2861–2881.

2. Geng C, Jung Y, Renaud N, Honavar V, Bonvin AMJJ, Xue LC: **iScore: A novel graph kernel-based function for scoring protein-protein docking models**. *Bioinformatics* 2019, **36**:112–121.

3. Renaud N, Geng C, Georgievska S, Ambrosetti F, Ridder L, Marzella D, Bonvin AMJJ, Xue LC: **DeepRank: A deep learning framework for data mining 3D protein-protein interfaces**. *Biorxiv* 2021, doi:10.1101/2021.01.29.425727.

4. Wang X, Terashi G, Christoffer CW, Zhu M, Kihara D: **Protein docking model evaluation by 3D deep convolutional neural networks**. *Bioinformatics* 2019, **36**:2113–2118.

5. Wang X, Flannery ST, Kihara D: **Protein Docking Model Evaluation by Graph Neural Networks**. *Biorxiv* 2020, doi:10.1101/2020.12.30.424859.

6. Gainza P, Sverrisson F, Monti F, Rodolà E, Bronstein M, Correia B: **Deciphering interaction fingerprints from protein molecular surfaces**. *Biorxiv* 2019, doi:10.1101/606202.

7. Roel-Touris J, Bonvin AMJJ, Jiménez-García B: **LightDock goes information-driven**. *Bioinformatics* 2019, **36**:950–952.

8. Padhorny D, Porter KA, Ignatov M, Alekseenko A, Beglov D, Kotelnikov S, Ashizawa R, Desta I, Alam N, Sun Z, et al.: **ClusPro in rounds 38 to 45 of CAPRI: Toward combining template-based methods with free docking**. *Proteins Struct Funct Bioinform* 2020, **88**:1082–1090.

• 9. Echartea MER, Ritchie DW, Beauchêne IC de: **Using restraints in EROS-DOCK improves model quality in pairwise and multicomponent protein docking**. *Proteins Struct Funct Bioinform* 2020, **88**:1121–1128.

> This software, based on an exhaustive real space coarse grain docking algorithm that uses quaternion pi-ball representations of the rotational space, is for the first time applied to protein docking. Further, the authors have extended the algorithm to support the use of restraints and enable multi-body docking.

10. Porter KA, Padhorny D, Desta I, Ignatov M, Beglov D, Kotelnikov S, Sun Z, Alekseenko A, Anishchenko I, Cong Q, et al.: **Template-based modeling by ClusPro in CASP13 and the potential for using co-evolutionary information in docking**. *Proteins Struct Funct Bioinform* 2019, **87**:1241–1248.

11. Trellet M, Zundert G van, Bonvin AMJJ: **Structural Bioinformatics, Methods and Protocols**. In *Methods in Molecular Biology*. . Springer US; 2020:145–162.

12. Saponaro A, Maione V, Bonvin A, Cantini F: **Understanding Docking Complexes of Macromolecules Using HADDOCK: The Synergy between Experimental Data and Computations**. *Bio-protocol* 2020, **10**:e3793.

13. Rosell M, Rodríguez-Lumbreras LA, Romero-Durana M, Jiménez-García B, Díaz L, Fernández-Recio J: **Integrative modeling of protein-protein interactions with pyDock for the new docking challenges**. *Proteins Struct Funct Bioinform* 2020, **88**:999–1008.

14. Rodrigues JPGLM, Bonvin AMJJ: **Integrative computational modeling of protein interactions**. *Febs J* 2014, **281**:1988–2003.

15. Saltzberg DJ, Viswanath S, Echeverria I, Chemmama IE, Webb B, Sali A: **Using Integrative Modeling Platform to compute, validate, and archive a model of a protein complex structure**. *Protein Sci* 2021, **30**:250–261.

16. Kim SJ, Fernandez-Martinez J, Nudelman I, Shi Y, Zhang W, Raveh B, Herricks T, Slaughter BD, Hogan JA, Upla P, et al.: **Integrative structure and functional anatomy of a nuclear pore complex**. *Nature* 2018, **555**:475–482.

17. Lasker K, Förster F, Bohn S, Walzthoeni T, Villa E, Unverdorben P, Beck F, Aebersold R, Sali A, Baumeister W: **Molecular architecture of the 26S proteasome holocomplex determined by an integrative approach**. *Proc National Acad Sci* 2012, **109**:1380–1387.

18. Viswanath S, Bonomi M, Kim SJ, Klenchin VA, Taylor KC, Yabut KC, Umbreit NT, Epps HAV, Meehl J, Jones MH, et al.: **The molecular architecture of the yeast spindle pole body core determined by Bayesian integrative modeling**. *Mol Biol Cell* 2017, **28**:3298–3314.

19. Engen JR, Botzanowski T, Peterle D, Georgescauld F, Wales TE: **Developments in Hydrogen/Deuterium Exchange Mass Spectrometry**. *Anal Chem* 2020, **93**:567–582.

20. Lite T-LV, Grant RA, Nocedal I, Littlehale ML, Guo MS, Laub MT: **Uncovering the basis of protein-protein interaction specificity with a combinatorially complete library**. *Elife* 2020, **9**:e60924.

21. Mintseris J, Gygi SP: **High-density chemical cross-linking for modeling protein interactions**. *Proc National Acad Sci* 2020, **117**:93–102.

22. Klykov O, Steigenberger B, Pektaş S, Fasci D, Heck AJR, Scheltema RA: **Efficient and robust proteome-wide approaches for cross-linking mass spectrometry**. *Nat Protoc* 2018, **13**:2964–2990.

23. Kalinin S, Peulen T, Sindbert S, Rothwell PJ, Berger S, Restle T, Goody RS, Gohlke H, Seidel CAM: **A toolkit and benchmark study for FRET-restrained high-precision structural modeling**. *Nat Methods* 2012, **9**:1218–1225.

24. Hopf TA, Green AG, Schubert B, Mersmann S, Schärfe CPI, Ingraham JB, Toth-Petroczy A, Brock K, Riesselman AJ, Palmedo P, et al.: **The EVcouplings Python framework for coevolutionary sequence analysis**. *Bioinformatics* 2018, **35**:1582–1584.

25. Dominguez C, Boelens R, Bonvin AMJJ: **HADDOCK: A Protein−Protein Docking Approach Based on Biochemical or Biophysical Information**. *J Am Chem Soc* 2003, **125**:1731–1737.

26. Vries SJ de, Schindler CE, Beauchene IC de, Zacharias M: **A web interface for easy flexible protein-protein docking with ATTRACT**. *Biophysical journal* 2015, **108**:462--465.

27. Pierce BG, Wiehe K, Hwang H, Kim B-H, Vreven T, Weng Z: **ZDOCK server: interactive docking prediction of protein--protein complexes and symmetric multimers**. *Bioinformatics* 2014, **30**:1771--1773.

28. Xia B, Vajda S, Kozakov D: **Accounting for pairwise distance restraints in FFT-based protein--protein docking**. *Bioinformatics* 2016, **32**:3342--3344.

29. Chelliah V, Blundell TL, Fernandez-Recio J: **Efficient restraints for protein--protein docking by comparison of observed amino acid substitution patterns with those predicted from local environment**. *Journal of molecular biology* 2006, **357**:1669--1682.

30. Kurcinski M, Jamroz M, Blaszczyk M, Kolinski A, Kmiecik S: **CABS-dock web server for the flexible docking of peptides to proteins without prior knowledge of the binding site**. *Nucleic acids research* 2015, **43**:W419--W424.

31. Kurcinski M, Ciemny MP, Oleniecki T, Kuriata A, Badaczewska-Dawid AE, Kolinski A, Kmiecik S: **CABS-dock standalone: a toolbox for flexible protein--peptide docking**. *Bioinformatics* 2019, **35**:4170--4172.

32. Jimenez-Garcia B, Roel-Touris J, Romero-Durana M, Vidal M, Jimenez-Gonzalez D, Fernandez-Recio J: **LightDock: a New Multi-Scale Approach to Protein-Protein Docking**. *Bioinformatics* 2018, **34**:49--55.

33. Echartea MER, Beauchêne IC de, Ritchie DW: **EROS-DOCK: protein–protein docking using exhaustive branch-and-bound rotational search**. *Bioinformatics* 2019, **35**:5003–5010.

34. Zacharias M: **ATTRACT: Protein–protein docking in CAPRI using a reduced protein model**. *Proteins Struct Funct Bioinform* 2005, **60**:252–256.

35. Hwang H, Vreven T, Janin J, Weng Z: **Protein–protein docking benchmark version 4.0**. *Proteins Struct Funct Bioinform* 2010, **78**:3111–3114.

36. Vries SJ de, Zacharias M: **ATTRACT-EM: a new method for the computational assembly of large molecular machines using cryo-EM maps**. *PLOS one* 2012, **7**:e49733.

37. Schindler CE, Vries SJ de, Sasse A, Zacharias M: **SAXS data alone can generate high-quality models of protein-protein complexes**. *Structure* 2016, **24**:1387--1397.

38. Viswanath S, Dominguez L, Foster LS, Straub JE, Elber R: **Extension of a protein docking algorithm to membranes and applications to amyloid precursor protein dimerization**. *Proteins: Structure, Function, and Bioinformatics* 2015, **83**:2170--2185.

39. Sali A, Blundell TL: **Comparative protein modelling by satisfaction of spatial restraints**. *Journal of molecular biology* 1993, **234**:779--815.

40. Schneidman-Duhovny D, Hammel M, Tainer JA, Sali A: **FoXS, FoXSDock and MultiFoXS: Single-state and multi-state structural modeling of proteins and their complexes based on SAXS profiles**. *Nucleic acids research* 2016, **44**:W424--W429.

41. Schneidman-Duhovny D, Hammel M, Sali A: **Macromolecular docking restrained by a small angle X-ray scattering profile**. *Journal of structural biology* 2011, **173**:461--471.

42. Zundert GC van, Melquiond AS, Bonvin AM: **Integrative modeling of biomolecular complexes: HADDOCKing with cryo-electron microscopy data**. *Structure* 2015, **23**:949--960.

43. Vries SJ de, Bonvin AM: **CPORT: a consensus interface predictor and its performance in prediction-driven docking with HADDOCK**. *PloS one* 2011, **6**:e17695.

44. Yu J, Vavrusa M, Andreani J, Rey J, Tufféry P, Guerois R: **InterEvDock: a docking server to predict the structure of protein–protein interactions using evolutionary information**. *Nucleic Acids Res* 2016, **44**:W542–W549.

45. Quignot C, Rey J, Yu J, Tufféry P, Guerois R, Andreani J: **InterEvDock2: an expanded server for protein docking using evolutionary and biological information from homology models and multimeric inputs**. *Nucleic Acids Res* 2018, **46**:gky377-.

• 46. Rudden LS, Degiacomi MT: **Transmembrane Protein Docking with JabberDock**. *Journal of Chemical Information and Modeling* 2021,

The authors tackle the modelling of transmembrane protein complexes docking, a category of interaction which has a low representation in the PDB database (4% of total structures) mostly due to its experimental determination challanges. JabberDock uses a shape representation of the membrane protein structure derived after short molecular dynamics simulations. On a self-made benchmark of 20 alpha-helix transmembrane helix proteins JabberDock achieves a success rate of 75%, the highest observed so far for transmembrane docking.

•• 47. Roel-Touris J, Jimenez-Garcia B, Bonvin AM: **Integrative modeling of membrane-associated protein assemblies**. *Nature communications* 2020, **11**:1--11.

Using LightDock, a flexible framework for the determination of protein complexes based on the Glowworm Swarm Optimisation algorithm, the authors describe a protocol that allow to account for the topological information provided by the membrane to guide the docking process. The resulting models are refined with HADDOCK to remove clashes. This work expands the capabilities of LightDock as integrative modelling software (see also reference 7).

48. Hurwitz N, Schneidman-Duhovny D, Wolfson HJ: **Memdock: an $\alpha$-helical membrane protein docking algorithm**. *Bioinformatics* 2016, **32**:2444--2450.

49. Jimenez-Garcia B, Pons C, Svergun DI, Bernado P, Fernandez-Recio J: **pyDockSAXS: protein--protein complex structure by SAXS and computational docking**. *Nucleic acids research* 2015, **43**:W356--W361.

50. Jeliazkov JR, Frick R, Zhou J, Gray JJ: **Robustification of RosettaAntibody and Rosetta SnugDock**. *Biorxiv* 2020, doi:10.1101/2020.05.26.116210.

51. Xia B, Mamonov A, Leysen S, Allen KN, Strelkov SV, Paschalidis IC, Vajda S, Kozakov D: **Accounting for observed small angle X-ray scattering profile in the protein--protein docking server cluspro**. *null* 2015,

52. Snderby P, Rinnan A, Madsen JJ, Harris P, Bukrinski JT, Peters GH: **Small-angle X-ray scattering data in combination with RosettaDock improves the docking energy landscape**. *Journal of chemical information and modeling* 2017, **57**:2463--2475.

53. Leman JK, Mueller BK, Gray JJ: **Expanding the toolkit for membrane protein modeling in Rosetta**. *Bioinformatics* 2017, **33**:754--756.

54. Ambrosetti F, Jiménez-García B, Roel-Touris J, Bonvin AMJJ: **Modeling Antibody-Antigen Complexes by Information-Driven Docking**. *Structure* 2020, **28**:119-129.e2.

55. Ambrosetti F, Jandova Z, Bonvin AMJJ: **A protocol for information-driven antibody-antigen modelling with the HADDOCK2.4 webserver**. *null* 2020,

• 56. Guest JD, Vreven T, Zhou J, Moal I, Jeliazkov JR, Gray JJ, Weng Z, Pierce BG: **An expanded benchmark for antibody-antigen docking and affinity prediction reveals insights into antibody recognition determinants**. *Structure* 2021, doi:10.1016/j.str.2021.01.005.

   The prediction of antibody-antigen complexes has been a challenge for the field of computational biology and its of great interest for the development of pharmaceuticals. In this publication an expanded benchmark is presented, containing more than double the amount of targets and binding affinities in comparison to previous benchmarks. The performance of several docking software and binding affinity predictors are compared.

57. Förster A, Schulze-Briese C: **A shared vision for macromolecular crystallography over the next five years**. *Struct Dynam-us* 2019, **6**:064302.

58. Kwan TOC, Axford D, Moraes I: **Membrane protein crystallography in the era of modern structural biology**. *Biochem Soc T* 2020, **48**:2505–2524.

59. Newport TD, Sansom MSP, Stansfeld PJ: **The MemProtMD database: a resource for membrane-embedded protein structures and their lipid interactions**. *Nucleic Acids Res* 2018, **47**:gky1047-.

60. Hopf TA, Schärfe CPI, Rodrigues JPGLM, Green AG, Kohlbacher O, Sander C, Bonvin AMJJ, Marks DS: **Sequence co-evolution gives 3D contacts and structures of protein complexes**. *Elife* 2014, **3**:e03430.

61. Zerihun MB, Pucci F, Peter EK, Schug A: **pydca v1.0: a comprehensive software for direct coupling analysis of RNA and protein sequences**. *Bioinformatics* 2019, **36**:2264–2265.

• 62. Nadaradjane AA, Quignot C, Traoré S, Andreani J, Guerois R: **Docking proteins and peptides under evolutionary constraints in Critical Assessment of PRediction of Interactions rounds 38 to 45**. *Proteins Struct Funct Bioinform* 2020, **88**:986–998.

   In this paper, the authors, using InterEvDock, nicely demonstrate the power of evolutionary data in CAPRI round 38-45. Using evolutionary constraints they were able to generate at least Acceptable models in their top 5 predictions in 11 (out of the total 16) targets, for 12 different interfaces with five Medium models and four High-quality models

63. Janin J: **Assessing predictions of protein–protein interaction: The CAPRI experiment**. *Protein Sci* 2005, **14**:278–283.

64. Rosell M, Fernández-Recio J: **Docking approaches for modeling multi-molecular assemblies**. *Curr Opin Struc Biol* 2020, **64**:59–65.

•• 65. Green AG, Elhabashy H, Brock KP, Maddamsetti R, Kohlbacher O, Marks DS: **Large-scale discovery of protein interactions at residue resolution using co-evolution calculated from genomic sequences**. *Nat Commun* 2021, **12**:1396.

    The authors use co-evolution for large-scale interaction prediction at residue resolution. They predict both the pairing of proteins, providing a probability score that two proteins interacts, and the contacts between them. Those contacts are then used in information-driven docking with HADDOCK to generate 3D models of the complexes. Using this approach they predict 504 interactions de novo in the E. coli membrane proteome.

66. Vajdi A, Zarringhalam K, Haspel N: **Patch-DCA: improved protein interface prediction by utilizing structural information and clustering DCA scores**. *Bioinformatics* 2019, doi:10.1093/bioinformatics/btz791.

67. Pazos F, Helmer-Citterich M, Ausiello G, Valencia A: **Correlated mutations contain information about protein-protein interaction 11Edited by A. R. Fersht**. *J Mol Biol* 1997, **271**:511–523.

68. Zeng H, Wang S, Zhou T, Zhao F, Li X, Wu Q, Xu J: **ComplexContact: a web server for inter-protein contact prediction using deep learning**. *Nucleic Acids Res* 2018, **46**:gky420-.

69. Shrestha R, Fajardo E, Gil N, Fidelis K, Kryshtafovych A, Monastyrskyy B, Fiser A: **Assessing the accuracy of contact predictions in CASP13**. *Proteins Struct Funct Bioinform* 2019, **87**:1058–1068.

70. Chacon P, Wriggers W: **Multi-resolution contour-based fitting of macromolecular structures**. *Journal of molecular biology* 2002, **317**:375--384.

71. Lasker K, Topf M, Sali A, Wolfson HJ: **Inferential optimization for simultaneous fitting of multiple components into a CryoEM map of their assembly**. *Journal of molecular biology* 2009, **388**:180--194.

72. Topf M, Lasker K, Webb B, Wolfson H, Chiu W, Sali A: **Protein structure fitting and refinement guided by cryo-EM density**. *Structure* 2008, **16**:295--307.

73. Putnam CD, Hammel M, Hura GL, Tainer JA: **X-ray solution scattering (SAXS) combined with crystallography and computation: defining accurate macromolecular structures, conformations and assemblies in solution**. *Quarterly reviews of biophysics* 2007, **40**:191--285.

74. Petoukhov MV, Svergun DI: **Global rigid body modeling of macromolecular complexes against small-angle scattering data**. *Biophysical journal* 2005, **89**:1237--1250.

75. Pons C, D'Abramo M, Svergun DI, Orozco M, Bernado P, Fernandez-Recio J: **Structural characterization of protein--protein complexes by integrating computational docking with small-angle scattering data**. *Journal of molecular biology* 2010, **403**:217--230.

76. Karaca E, Bonvin AM: **On the usefulness of ion-mobility mass spectrometry and SAXS data in scoring docking decoys**. *Acta Crystallographica Section D: Biological Crystallography* 2013, **69**:683--694.

77. Franke D, Petoukhov M, Konarev P, Panjkovich A, Tuukkanen A, Mertens H, Kikhney A, Hajizadeh N, Franklin J, Jeffries C, et al.: **ATSAS 2.8: a comprehensive data analysis suite for small-angle scattering from macromolecular solutions**. *Journal of applied crystallography* 2017, **50**:1212--1225.

78. Manalastas-Cantos K, Konarev PV, Hajizadeh NR, Kikhney AG, Petoukhov MV, Molodenskiy DS, Panjkovich A, Mertens HD, Gruzinov A, Borges C, et al.: **ATSAS 3.0: expanded functionality and new tools for small-angle scattering data analysis**. *Journal of Applied Crystallography* 2021, **54**.

79. Russel D, Lasker K, Webb B, Velázquez-Muriel J, Tjioe E, Schneidman-Duhovny D, Peterson B, Sali A: **Putting the Pieces Together: Integrative Modeling Platform Software for Struct**